# Dosimetric calibration of an anatomically specific ultra-high dose rate electron irradiation platform for preclinical FLASH radiobiology experiments

Jinghui Wang, PhD[1,6*], Stavros Melemenidis, PhD[1*], Rakesh Manjappa, PhD[1], Vignesh Viswanathan PhD[1], Ramish M. Ashraf, PhD[1], Karen Levy, MD[3], Lawrie Skinner, PhD[1], Luis A. Soto PhD[1], Stephanie Chow, MD[3], Brianna Lau[1], Ryan B. Ko[1], Edward E. Graves, PhD[1,2], Amy S. Yu, PhD[1], Karl K. Bush, PhD[1,7], Murat Surucu, PhD[1], Erinn B. Rankin, PhD[1,2,3], Billy W. Loo Jr., MD, PhD[1,2], Emil Schüler, PhD[4#], Peter G. Maxim, PhD[5#]

[1] Department of Radiation Oncology, Stanford University School of Medicine, Stanford, CA 94305, USA.
[2] Stanford Cancer Institute, Stanford University School of Medicine, Stanford, CA 94305, USA.
[3] Department of Gynecologic Oncology, Stanford University School of Medicine, Stanford, CA, 94305, USA.
[4] Department of Radiation Physics, Division of Radiation Oncology, The University of Texas MD Anderson Cancer Center, Houston, TX, 77030, USA.
[5] Department of Radiation Oncology, University of California Irvine School of Medicine, Orange, CA 92868, USA.
[6] Guangdong Institute of Laser Plasma Accelerator Technology, Guangzhou, Guangdong, 510080, China.
[7] Varian Medical Systems, Palo Alto, CA 94304, USA.

# Author contributions

All authors met the International Committee of Medical Journal Editors (ICMJE) criteria for authorship (https://www.icmje.org/recommendations/browse/roles-and-responsibilities/defining-the-role-of-authors-and-contributors.html). Jinghui Wang and Stavros Melemenidis are co-first authors, each of whom are appropriately listed first when referencing this publication on their respective CVs. Emil Schüler and Peter G. Maxim, PhD are co-senior/co-corresponding authors.

#**Corresponding authors:**
Emil Schüler, PhD
*Department of Radiation Physics, Division of Radiation Oncology, The University of Texas MD Anderson Cancer Center, Houston, 77030 TX, USA Email: eschueler@mdanderson.org*

Peter G. Maxim, PhD
*Department of Radiation Oncology, University of California Irvine, Irvine, CA 92698, USA*
*Email: pmaxim@uci.edu*

*****Co-first authors:**
Jinghui Wang, PhD
*Current address, Guangdong Institute of Laser Plasma Accelerator Technology, Guangzhou, Guangdong, 510080, China.*
*Email: wangjinghui@glapa.cn*

Stavros Melemenidis, PhD
*Department of Radiation Oncology, Stanford University School of Medicine, Stanford, CA 94305, US*
*Email: stavmel@stanford.edu*

**Author for editorial correspondence:**
Stavros Melemenidis, PhD
*Department of Radiation Oncology, Stanford University School of Medicine, Stanford, CA 94305, US*
*Email: stavmel@stanford.edu*

**Statistical analysis**
Jinghui Wang, PhD
*Guandong Institute of Laser Plasma Accelerator Technology, Guangzhou, Guangdong, 510080, China.*
*Email: wangjinghui@glapa.cn*

# Funding statement

This work was supported by the Office of the Assistant Secretary of Defense for Health Affairs

through the Department of Defense Ovarian Cancer Research Program under Award No. W81XWH-17-1-0042; the My Blue Dots fund; the Stanford University Department of Radiation Oncology; the Weston Havens Foundation; the Stanford University School of Medicine; the Stanford University Office of the Provost; the Wallace H. Coulter Foundation; the Cancer League; the Swedish Childhood Cancer Foundation; the Foundation BLANCEFLOR Boncompagni Ludovisi n'ee Bildt; the American Association for Cancer Research; National Institutes of Health P01 CA244091 (EEG, PGM, BWL), R01 CA26667 (ES, PGM, EEG, BWL), R01CA233958 (BWL); NIH Cancer Center Support (Core) Grant P30 CA016672 (ES) from the National Cancer Institute, National Institutes of Health to The University of Texas MD Anderson. We also gratefully acknowledge philanthropic donors to the Department of Radiation Oncology at Stanford University School of Medicine.

**Disclosures**

PGM and BWL have received research support from Varian Medical Systems. PGM and BWL are co-founders of TibaRay. BWL is a board member of TibaRay. BWL is a consultant on a clinical trial steering committee for Beigene and has received lecture honoraria from Mevion. KKB is currently an employee of Varian Medical Systems. All other authors declare no conflicts of interest.

**Acknowledgements**

We thank Christine F. Wogan, MS, ELS, of MD Anderson's Division of Radiation Oncology, for editorial contributions to this article.

# Short running title:

Clinical Linac Configuration for Preclinical FLASH

# Abstract

**Purpose:** We characterized the dosimetric properties of a clinical linear accelerator configured to deliver ultra-high dose rate (UHDR) irradiation to two anatomic sites in mice and for cell-culture FLASH radiobiology experiments.

**Methods:** Delivered doses of UHDR electron beams were controlled by a microcontroller and relay interfaced with the respiratory gating system. We also produced beam collimators with indexed stereotactic mouse positioning devices to provide anatomically specific preclinical treatments. Treatment delivery was monitored directly with an ionization chamber, and charge measurements were correlated with radiochromic film measurements at the entry surface of the mice. The setup for conventional (CONV) dose rate irradiation was similar but the source-to-surface distance was longer. Monte Carlo simulations and film dosimetry were used to characterize beam properties and dose distributions.

**Results:** The mean electron beam energies before the flattening filter were 18.8 MeV (UHDR) and 17.7 MeV (CONV), with corresponding values at the mouse surface of 17.2 MeV and 16.2 MeV. The charges measured with an external ion chamber were linearly correlated with the mouse entrance dose. Use of relay gating for pulse control initially led to a delivery failure rate of 20% (±1 pulse); adjustments to account for the linac latency improved this rate to <1/20. Beam field sizes for two anatomically specific mouse collimators (4×4 $cm^2$ for whole-abdomen and 1.5×1.5 $cm^2$ for unilateral lung irradiation) were accurate within <5% and had low radiation leakage (<4%). Normalizing the dose at the center of the mouse (~0.75 cm depth) produced UHDR and CONV doses to the irradiated volumes with >95% agreement.

**Conclusions:** We successfully configured a clinical linear accelerator for increased output and developed a robust preclinical platform for anatomically specific irradiation, with highly accurate and precise temporal and spatial dose delivery, for both CONV and UHDR applications.

**Keywords:** FLASH, ultra-high dose rate, radiotherapy, dosimetry, Linac

# Introduction

The main limitation of curative cancer radiation therapy (RT) is toxicity to healthy tissues that are unavoidably exposed along the beams that target the tumor. Ultra-high dose rate (UHDR) irradiation (>40 Gy/s) causes less toxicity to healthy tissues than conventional (CONV) dose rate irradiation (~0.1 Gy/s) but does not compromise tumor control[1,2]. This radiobiological sparing effect of UHDR on normal tissues, the FLASH effect, has a reported dose-modifying factor of 1.1–1.4[3] and is strongly influenced by dose, fractionation schedule, tissue, and endpoint under investigation. The potential for widening the therapeutic window by translating FLASH to clinical settings has ignited great interest in the research community. From an institutional standpoint, investigating the radiobiological mechanism(s) behind the FLASH effect and preparing for clinical applications have focused on approaches to generate UHDR irradiation[4], which to date have included configuring X-ray tubes[5,6], synchrotron microbeams[7], proton facilities[8–10], laser-driven electron beams[11], linear induction accelerators[12], and clinical linear accelerators[13] to generate UHDR treatments in preclinical settings[14]. A clinical FLASH project involving conformal UHDR MV X-rays, 'Pluridirectional High-energy Agile Scanning Electronic Radiotherapy (PHASER),'' is currently ongoing[15], and other electron FLASH platforms are under development[16]. The widespread availability of clinical linear accelerators (linacs) makes configuring them for early FLASH clinical investigations a practical approach[17,18].

The emergence of FLASH poses scientific and technical challenges. The effects of beam characteristics (e.g., instantaneous dose rate, pulse duration, duty cycle, and pulse frequency) on the FLASH effect have yet to be determined, and the radiobiological mechanism underlying the FLASH effect is not well understood[2,19]. Efforts are ongoing to design (i) collimators to deliver accurate spatial dose distribution; (ii) ion chambers to overcome charge saturation to measure absolute dose; and (iii) systems for counting and delivering accurate numbers of pulses. Meeting the increasing demands for radiobiological research and the rapid pace of technologic advances for next-generation RT systems can be facilitated by configuring additional clinical linacs with FLASH capability[20].

At Stanford Cancer Center, we created a previous FLASH platform by configuring the decommissioned 20-MeV energy program board of a Varian Clinac 21 EX system[18] and used that platform for several biological studies[21–24]. However, that platform was decommissioned in 2018. For this report, we configured a Varian Trilogy clinical linac to deliver UHDR electron beams. Some of the beam properties of this platform at long source-to-surface distances (SSDs) have been measured and reported[25]. Here we describe technical advances in using this device for *in vitro* and *in vivo* preclinical studies, including how to confine the radiation field, deliver accurate numbers of pulses, and measure the dose. We also characterized and compared dosimetric characteristics of UHDR and CONV setups, including electron energies, beam sizes, and achievable dose rates at various distances from the radiation source (the scattering foil).

## Materials and Methods

### Configuring a Linac for UHDR

A Varian Trilogy clinical RT system (Varian Medical Systems, Palo Alto, CA) at Stanford Cancer Center was configured to deliver a UHDR electron beam as follows. The 20-MeV electron program circuit board was decommissioned from clinical practice to allow the circuit board to be tuned and the beam energy to be adjusted and reduced to match the 16-MeV clinical electron beam. The bending magnet current, the gun's high voltage, and the gun grid voltage were all tuned and optimized for beam output. The corresponding electron flattening filter was replaced with a filter used for the 16-MeV energy electron beam. The **Supplemental Material** section **'Operation'** describes the settings associated with the control room.

**Beam control**

The respiratory gating control interface of the linac is connected to a microcontroller (Red Pitaya, Slovenia, Europe) that hosts a mechanical relay circuit. In this configuration, the machine can be in "beam on" status but the pulses are not fired. The pulse signal from the linac's ionization chamber (located after the gun; 'top TP1') is fed to the microcontroller, which calculates the time required to deliver the prescribed number of pulses. The microcontroller closes the relay to initiate pulses and opens the relay at the end of the desired time window to stop the delivery of pulses through the linac's gating circuit (i.e., relay gating; **Fig 1A,B**). A detailed description of this configuration can be found in the **Supplemental Material** section **'Pulse control and pulse counting'**.

**Mouse irradiations**

All animal experiments and procedures were approved by the Institutional Animal Care and Use Committee of Stanford University and were conducted in accordance with institutional and NIH guidelines (APLAC 27939). A total of 32 eight- to ten-week-old C57BL/6J female mice were purchased from Jackson Laboratories (Sacramento, CA). Mice were fully anaesthetized before irradiation by intraperitoneal injection of a mixture of ketamine (100 mg/kg) and xylazine (10 mg/kg).

      Each organ-specific irradiation includes a dedicated beam collimator with the appropriate

exposure field aperture (**Supplemental Fig. S1A**). **Figure 1C** shows an anesthetized mouse positioned inside a whole-abdomen stereotactic mouse positioner (top) and illustrates the positioning of the radiochromic film used for dosimetry (bottom). The stereotactic mouse positioner is anchored on the top of the whole-abdomen collimator (**Fig. 1D**), and the two structures are inserted together at the bottom of a cradle as shown in **Figure 1E**. The top of the cradle is designed to fit precisely on the surface of the linac's treatment head, below the multileaf collimator, when the X and Y jaws are open at maximum (**Fig. 1F**). In-depth descriptions of the materials, fabrication, and method of stereotactic mouse positioning for organ-specific irradiation are supplied in the **Supplemental Material** sections **'Collimator design and production'**, **'Anatomically specific mouse collimators', 'Stereotactic mouse positioner',** and '**Cradle-collimator positioning.'**

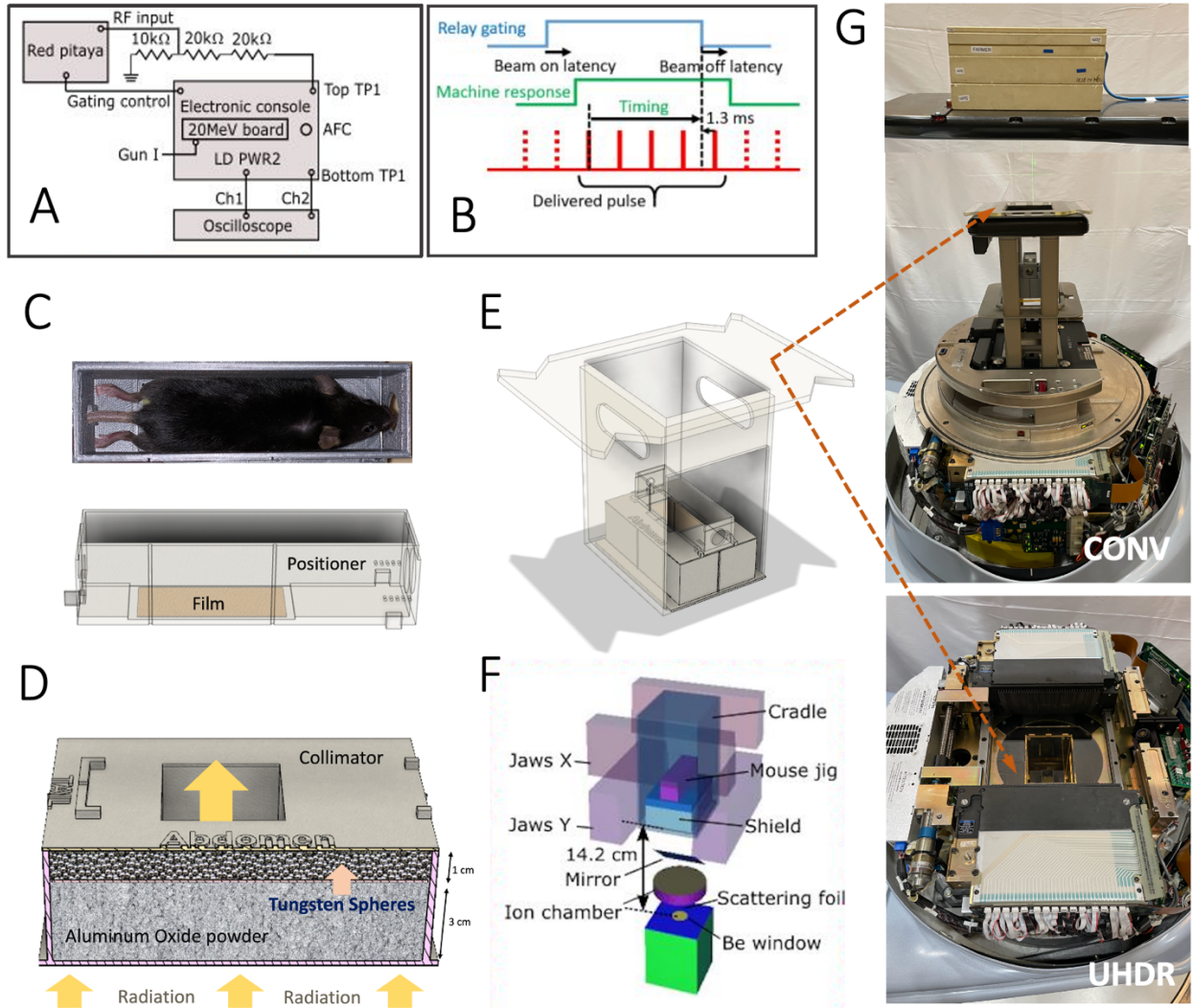

**Figure 1.** (A) Schematic of the electronic circuit that controls pulse delivery during ultra-high dose rate (UHDR) irradiations. 'Red Pitaya' is the manufacturer of the microcontroller; TP1 indicates the ionization chamber; Gun I, gun current; AFC, automatic frequency controller. (B) Schematic of the relay pulse gating example, showing the duty circle (90 Hz), the linac's inherent latency, and the relay gating to deliver five pulses. (C) Top, the stereotactic mouse positioner used for whole-abdomen irradiation illustrating the positioning of an anesthetized mouse. Below, a 3D computer-aided design (CAD) file of the positioner showing the positioning of the radiochromic film used for dosimetry. (D) A 3D CAD file of the whole-abdomen collimator (4×4 cm) with a lateral cross-section, showing the shielding layers of low-Z and high-Z material. (E) A 3D CAD file of the assembled cradle with the whole-abdomen collimator and stereotactic mouse positioner. (F) For UHDR, the cradle can be placed inside the treatment head when the jaws are fully open, and the bottom of the shield is suspended at 14.2 cm from the scattering foil. (G) The UHDR and conventional (CONV) dose rate irradiation setup, where the assembled cradle is mounted either on the 15×15 cm electron applicator (top) or inside the treatment head (below). In both the UHDR and CONV geometries, an ion chamber and solid water are supported by the couch and allow pulse monitoring, as well as 10 cm of solid water build-up, farmer chamber solid water block, and 3-cm solid water blocks for backscatter.

### *In vivo* mouse irradiation geometries

The experimental setups for UHDR and CONV mouse irradiations *in vivo* are shown in **Figures 1G** and **2A**. The treatment head is rotated to 180 degrees (with the beam pointed at the ceiling) and the jaws X and Y are fully open for both modalities. For UHDR irradiation, the headcover is removed and the cradle with the collimator and stereotactic mouse positioner is mounted inside the treatment head. In this setup, the collimator's bottom surface is suspended at 14.2 cm from the scattering foil (radiation source), and the mouse entrance surface is suspended at 18.7 cm (source-to-film distance). For the CONV irradiation, a 15×15 cm$^2$ electron applicator is used. Achieving low dose rates requires positioning the mice such that the bottom surface of the collimator is suspended at 71.6 cm from the scattering foil, i.e., the source-to-film distance is 76.1 cm (**Fig. 2A,B**).

For both UHDR and CONV modalities, pulses are monitored with a PWT Farmer Chamber 30013 (Booton, NJ) positioned in the middle of the field and inside a solid water block structure. The setup is built once and remains unchanged for both UHDR and CONV modalities (**Fig. 1G, Fig. 2A**). The ion chamber allows measurement of the exit charge (through the mouse body), which is monitored from the electrometer in the control room. The beam monitoring methods are described in further detail in the **Supplemental Material** section **'Beam monitor for *in vivo* UHDR irradiation.'**

**Film dosimetry**

While the mice are being irradiated, a 2.4×5.1 cm$^2$ piece of EBT3 Gafchromic film (Ashland Inc., Wayne, NJ, USA) is loaded onto the stereotactic mouse positioner to measure the entrance surface absorbed dose. Films were also analyzed for the lateral profiles or percentage depth dose (PDD) by choosing a rectangular area of interest on the scanned datasets and averaging the pixels along the X or Y directions. Details of the film dosimetry are provided in the **Supplemental Material** section **'Film analysis.'**

**Derivation of electron beam energy**

To derive the beam energies at the entrance surface of the mice for both UHDR and CONV modalities, we used the FLUKA/FLAIR[26] Monte Carlo (MC) code to simulate the electron beam parameters. To verify the accuracy of the simulation, the MC results were compared with the analyzed PDDs and the lateral profiles from the irradiated films (as described in the previous section). Once agreement was established and the primary electron beam energy was identified, an "area scorer" of 2×2 cm$^2$ was set at 18.7 cm (UHDR) or 76.1 cm (CONV) from the scattering foil, and the electron energy spectra were derived (**Fig. 2B**). MC simulations were also used to determine the lateral dose distribution of each collimator at the radiation entrance point by using a 0.1×0.1×0.3 cm$^3$ voxel scorer, and associated PDDs by using a cylinder scorer (radius 0.2 cm, height 0.2 cm). For the open-field lateral dose distributions at different SSDs, concentric ring mesh scorers (0.3 cm high, 0.3 cm wide) were used. The number of primary electrons was chosen to keep the statistical uncertainty below 2% within the full width-half maximum (FWHM) region of the beam profile and whole PPD. The final step involved comparing the MC results with those from the radiochromic films.

**Characterizing field sizes for each anatomically specific mouse collimator**

Each anatomic mouse collimator has a different radiation field size (**Supplemental Fig S1A**) and must be calibrated individually for the entry dose for both UHDR and CONV geometries. Each collimator must also be characterized for electron beam homogeneity at the dose entrance in the X and Y directions (beam profile) and attenuation of the electron beam through the mouse body in the Z direction (PDD). For UHDR, each anatomic mouse collimator and its associated stereotactic mouse positioner were loaded with radiochromic film and 2-cm solid water phantom for backscatter to mimic maximum mouse height. The entrance surface absorbed dose from films was correlated with the exit charges from the ion chamber. For the PDDs, vertical films (2.4×5.1 $cm^2$, along the Z direction) were sandwiched in solid water blocks and placed inside the stereotactic mouse positioner. Three irradiations per condition were performed, and the measurements were averaged. In this report, only two mouse anatomic collimators were used: one with a large irradiation field (whole abdomen, 4×4 $cm^2$) and one with a smaller irradiation field (unilateral lung, 1.5×1.5 $cm^2$). For UHDR, the collimators were irradiated at different gun currents (Gun I) with a single pulse and 2-cm-thick solid water phantom (no film). The Gun I can be altered from a potentiometer located on the energy circuit board and can be read with a voltmeter. The Gun I was set for a range of 6 V to 14 V (in 1-V steps), and exit charge measurements from the ion chamber were correlated. With the Gun I voltage set at ~10 V, the collimators were irradiated with 2, 4, 8, 12, and 16 pulses (2-cm solid water phantom with film at the entrance surface) to derive the charge per Gy at the entrance surface. Notably, the collimators needed to be evaluated at only a single Gun I setting (at one dose per pulse) because the film is dose-rate-independent, and the energy of the beam does not change with Gun I setting because the settings of the bending magnet are held constant.

**Characterizing open fields for irradiating cell cultures**

For cell culture irradiation, typically flasks, dishes, or well plates are used and require a larger radiation field and a relatively flat dose distribution. To establish a radiation field area large enough for an UHDR electron beam, we used an open field without collimation. To characterize the beam profile, large radiochromic films (10×13 cm$^2$) were irradiated while supported perpendicular to the beam at various distances from the scattering foil (14.2, 19.2, 24.2, 29.2, 34.2, and 39.2 cm SSD) by using five 3D-printed film positioners (5 cm high) mounted on the cradle, stacked on top of each other (**Supplemental Fig. S1D).** The lateral profiles of the irradiated films were analyzed (**Supplemental Fig. S1E**), and the flatness of the beam was characterized as the maximum dose variation within the area of interest, normalized to the dose at the center of the field. The minimum radius between the negative and positive sides of the measured profile was used for analysis. For each SSD assessed, the relative dose rate was calculated. Once the SSD with the most useful flatness and dose rate was established (29.2 cm SSD), a cell culture holder was designed and printed to fit the cradle and to suspend flasks, dishes, or well plates at this distance.

# RESULTS

**Derivation of electron beam energy**

The PDDs from the films at the maximum possible elevation of the couch (i.e., SSD of 136.5 cm from the scattering foil) were analyzed for both UHDR and CONV beams (**Fig. 2B,C**). The projected ranges (Rp) for the UHDR and CONV electron beam were 8.2 and 7.7 cm, respectively. This confirms that the bremsstrahlung tail extended beyond 9 cm into solid water for both beams, and the ion chamber was thus placed at 147.5 cm with a 11.0-cm solid water build-up at the top of the couch (**Fig. 2A**).

The measured PDDs were simulated with a MC code, and the primary electron beam was tuned to match the measured PDD curves (**Fig. 2C**). The primary electron source before the scattering foil was calculated to have a mean energy of 18.8 MeV for UHDR and 17.7 MeV for CONV. These primary electron energies in MC simulations were used to derive the energy spectra of the electron beams for both UHDR and CONV geometries (**Fig. 2D**) at the distance to where the beam enters the mouse body (**Fig. 2A**). The mean energies for the UHDR and CONV beams were 17.2 MeV and 16.2 MeV, and the most probable energies were 18.1 MeV and 17.1 MeV (**Fig. 2D**). The electron energy for UHDR was about 1 MeV higher than that for CONV.

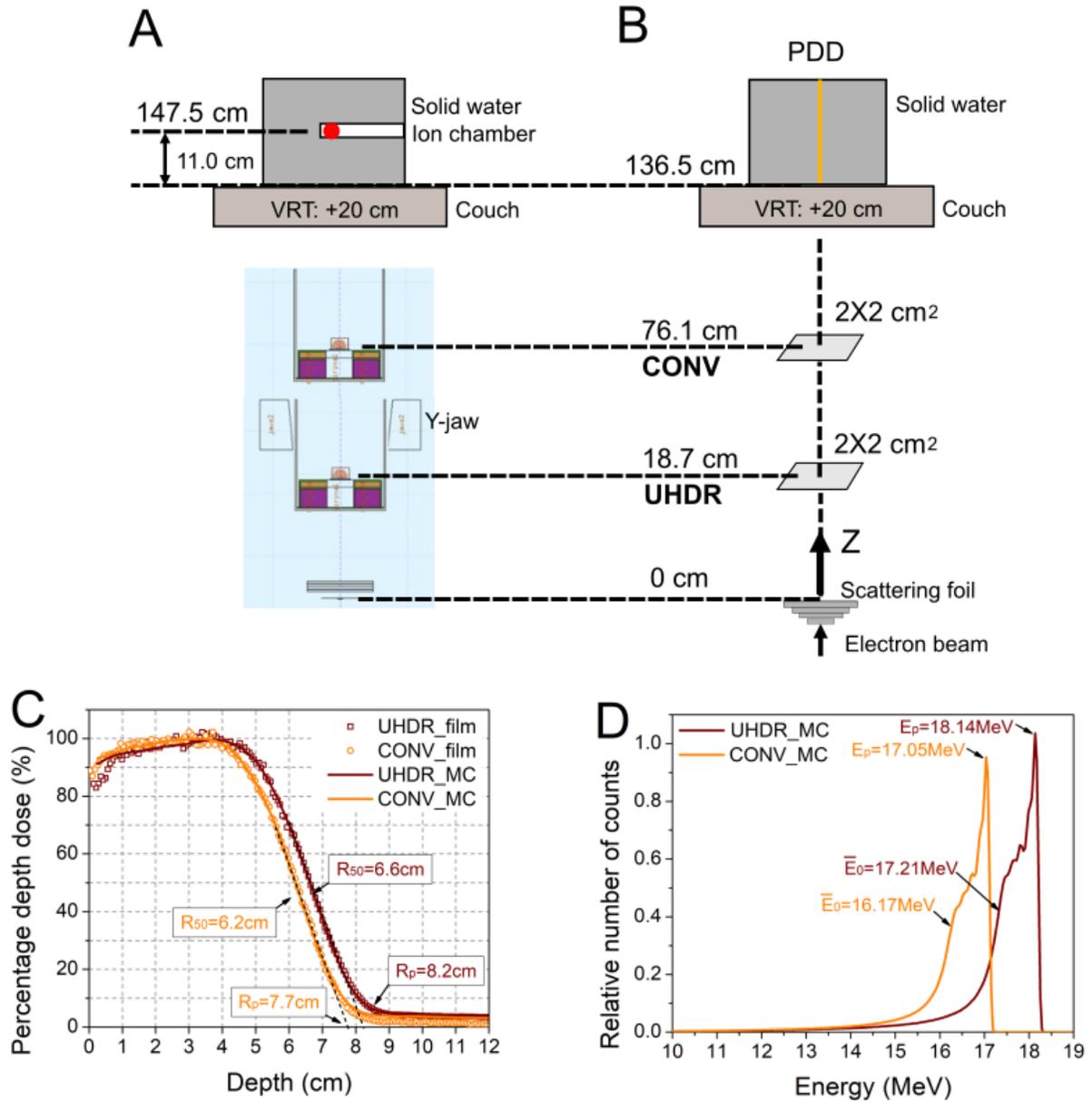

**Figure 2.** (A) Schematic illustrating the different geometries for *in vivo* ultra-high dose rate (UHDR) and conventional (CONV) dose rate irradiations. The distances from the scattering foil are indicated for the film placement, which is used to measure the absorbed entrance dose, and the farmer ion chamber used to measure the exit charge. VRT indicates the vertical position of the couch. (B) Schematic illustrating the Monte Carlo simulation of the experimental setting for the percentage depth doses (PDDs) and the simulation used to derive the energies at the height of the entrance of the beam into the mouse. (C) Film-measured Monte Carlo-simulated PDD curves on top of the couch (136.5 cm from the scattering foil) for both UHDR and CONV dose rate electron beams. (D) Simulated electron beam spectra for UHDR and CONV geometries at the height of the beam entrance into the mice, where the film is typically placed for dosimetry measurements. Detector size: 2×2 cm on the central axis. UHDR electron beam energy was ~1 MeV higher than CONV.

**Beam monitor for *in vivo* UHDR irradiation**

For UHDR irradiations, the dose per pulse is determined from fine-tuning the grid voltage (i.e., changing the voltage setting of Gun I) and is specific to the irradiation field size of each collimator. The relationship between the Gun I voltage setting and the exit charge, measured from a single pulse to the whole-abdomen and unilateral lung collimators (with a 2-cm solid water phantom inside each) is shown in **Figure 3A**. To derive the exit charge per film dose (pC/Gy) values, we irradiated the same collimator, with a phantom configuration and at 10 V grid voltage, with 2 to 22 pulses. The linear relationship between the film entrance dose and the exit charge for the two collimators, from which the pC/Gy can be derived, is shown in **Figure 3B**. The difference in the slope of pC/Gy between collimators is due to the radiation field size, with the whole-abdomen radiation field area being roughly 7 times larger than that of the unilateral lung. Therefore, the maximum possible doses per pulse were 4.1 Gy/pulse for the whole-abdomen collimator and 4.8 Gy/pulse for the unilateral lung collimator.

**Mouse irradiations**

For the whole-abdomen irradiations, 15 mice were treated with a 14-Gy entrance surface dose, with 7 pulses of 2 Gy/pulse, and an expected exit charge of 236.8 pC. Entrance doses and exit charges for the irradiated mice are shown in **Figure 3C**. The markings for the charges indicate that two mice received 6 of 7 pulses and one mouse received 8 pulses (for a failure rate of 3 of 15 [20%]). The mean entrance dose for the remaining 12 mice was 13.8 ± 0.3 Gy, and the mean exit charge was 241.1 ± 1.9 pC. The average deviation of the film dose from the target dose was 1.7%, and the average deviation of the exit charge measured with the ion chamber from the target charge was 1.8%. The dose rate was 210.0 Gy/s, and the instantaneous dose rate was 5.3E+5 Gy/s (3.75 µs pulse width).

For the unilateral lung irradiations, 17 mice were treated with surface entrance doses of 20 Gy (7 mice), 30 Gy (6 mice), or 40 Gy (4 mice), with 10, 15, or 20 pulses of 2 Gy/pulse and corresponding expected exit charges of 275.0 pC, 413.1 pC, and 551.3 pC (**Fig. 3D**). Again, the number of pulses unsuccessfully delivered can be easily identified from the figure (3 of 17 [18%]). The mean entrance dose for the successful deliveries was $20.1 \pm 0.2$ Gy for mice given 20 Gy, $29.6 \pm 0.6$ Gy for mice given 30 Gy, and $40.3 \pm 0.4$ Gy for mice given 40 Gy; the corresponding mean exit charges were $276.8 \pm 1.0$ pC, $414.6 \pm 1.9$ pC, and $551.2 \pm 3.3$ pC. The average deviations of the film dose from the target dose for each group were 0.5%, 2.1%, and 1.2%, and the average deviations of the exit charge measured with the ion chamber from the target charge for each group were 0.7%, 0.4%, and 0.4%. The dose rates were 200 Gy/s for the 20-Gy group, 192.9 Gy/s for the 30-Gy group, and 189.5 Gy/s for the 40-Gy group.

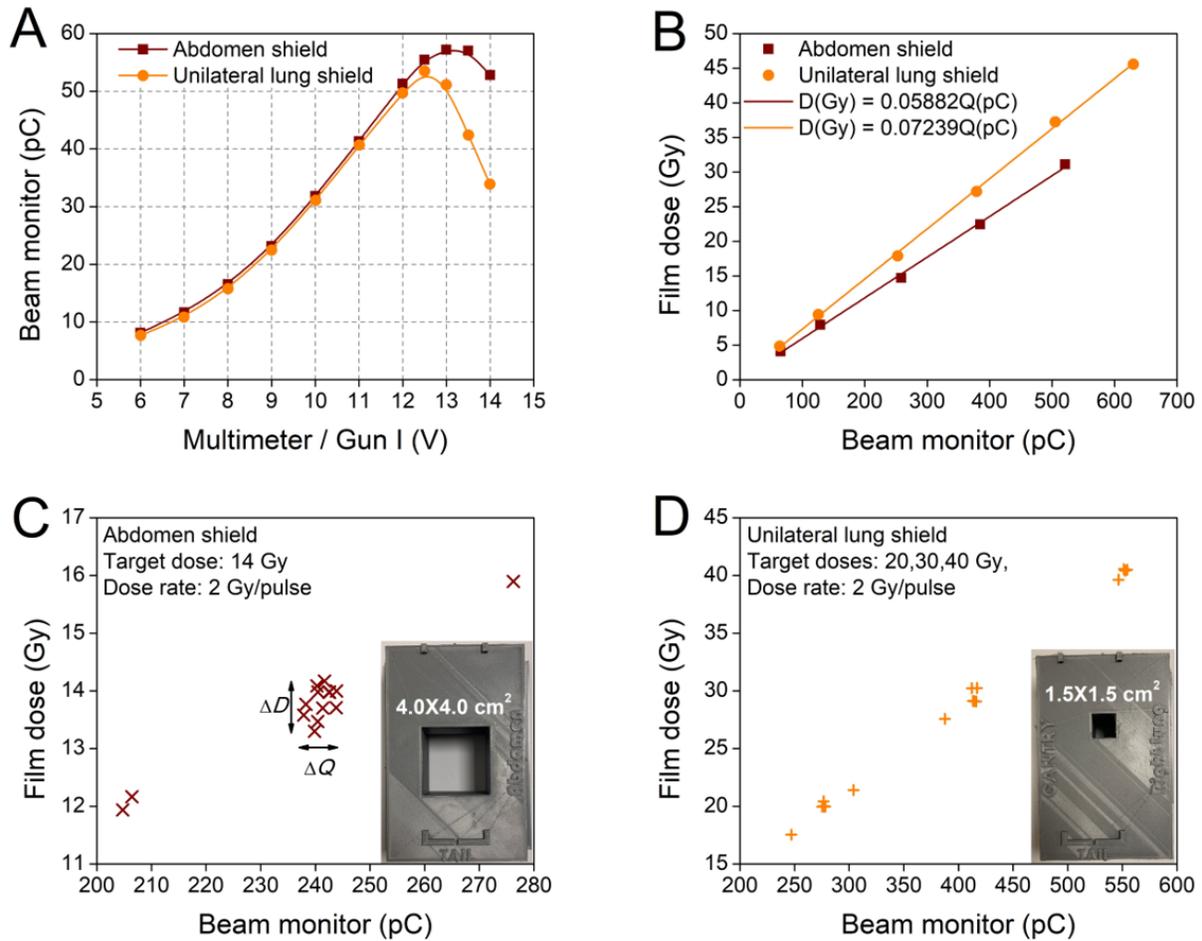

**Figure 3.** (A) Plot showing the beam monitor signal as a function of the linac's electron gun grid voltage (Gun I) for the whole-abdomen and unilateral lung irradiation mouse collimators, with a 2-cm solid water phantom inside the stereotactic mouse positioner. (B) Plot showing the linear relationship of film entrance dose and the exit charge for the whole-abdomen and unilateral lung irradiation mouse collimators, with a 2-cm solid water phantom and film inside the stereotactic mouse positioner. The films were irradiated with 2 to 22 pulses at a Gun I multimeter (grid voltage) reading of 10 V. (C, D) Film entrance doses and exit charges (with uncertainties indicated as ΔD or ΔQ) from in vivo UHDR mouse irradiations of (C) whole abdomen and (D) unilateral lung with mouse collimators. The irradiations with an extra pulse or a missing pulse (and hence with the delivered dose deviating from the targeted dose) can be easily identified from the beam monitoring readings. Dose uncertainty for the EBT3 film is 4%.

## Fields for anatomically specific mouse collimators

Film-derived transverse beam profiles at the point of entry in the mouse and the central PDD curves for whole-abdomen and unilateral lung collimators for both UHDR and CONV geometries, overlaid with MC simulations of the same measurements, are shown in **Figure 4A-D**. The MC simulations and experimental measurements were in excellent agreement; beam characteristics for both collimators (field size [FWHM], 20-80% penumbra, radiation leakage at 1 cm from the field edge, and depth dose attenuation) are summarized in **Table 1**.

**Table 1**. Beam characteristics for whole-abdomen and unilateral lung irradiation collimators derived from irradiated radiochromic films along the X, Y and Z orientation along the path of the beam.

|  | Whole Abdomen ($2\times2$ cm$^2$) | | Unilateral Lung ($1.5\times1.5$ cm$^2$) | |
| --- | --- | --- | --- | --- |
|  | X-axis, UHDR / CONV | Y-axis, UHDR / CONV | X-axis, UHDR / CONV | Y-axis, UHDR / CONV |
| **Beam profile FWHM, cm** | 2.07 / 2.02 | 2.06 / 2.00 | 1.57 / 1.53 | 1.57 / 1.53 |
| **20%-80% penumbra, cm*** | 0.14 / 0.30 | 0.15 / 0.29 | 0.28 / 0.41 | 0.23 / 0.37 |
| **Leakage, %** (1 cm from field edge **) | 2.9 / 4.0 | 1.8 / 3.2 | 2.0 / 3.7 | 1.5 / 3.9 |
| **Depth dose at depths, %** | | | | |
| 0.5 cm | 95 / 99 | | 93 / 92 | |
| 1.0 cm | 91 / 95 | | 86 / 88 | |
| 1.5 cm | 87 / 93 | | 79 / 76 | |
| 2.0 cm | 84 / 88 | | 69 / 67 | |

Abbreviations: UHDR, ultrahigh dose rate; CONV, conventional dose rate; FWHM, full width half maximum.

*Penumbras are calculated from the average distance separating the 20% and 80% isodose lines of the beam profile, and presented as a single average value of left and right penumbras of the beam profile.

**Leakage is measured from the film dose of the collimator's profile, chosen outside of the radiation field and calculated as a percentage of the central axis dose. Only the maximum between the left and right side is presented.

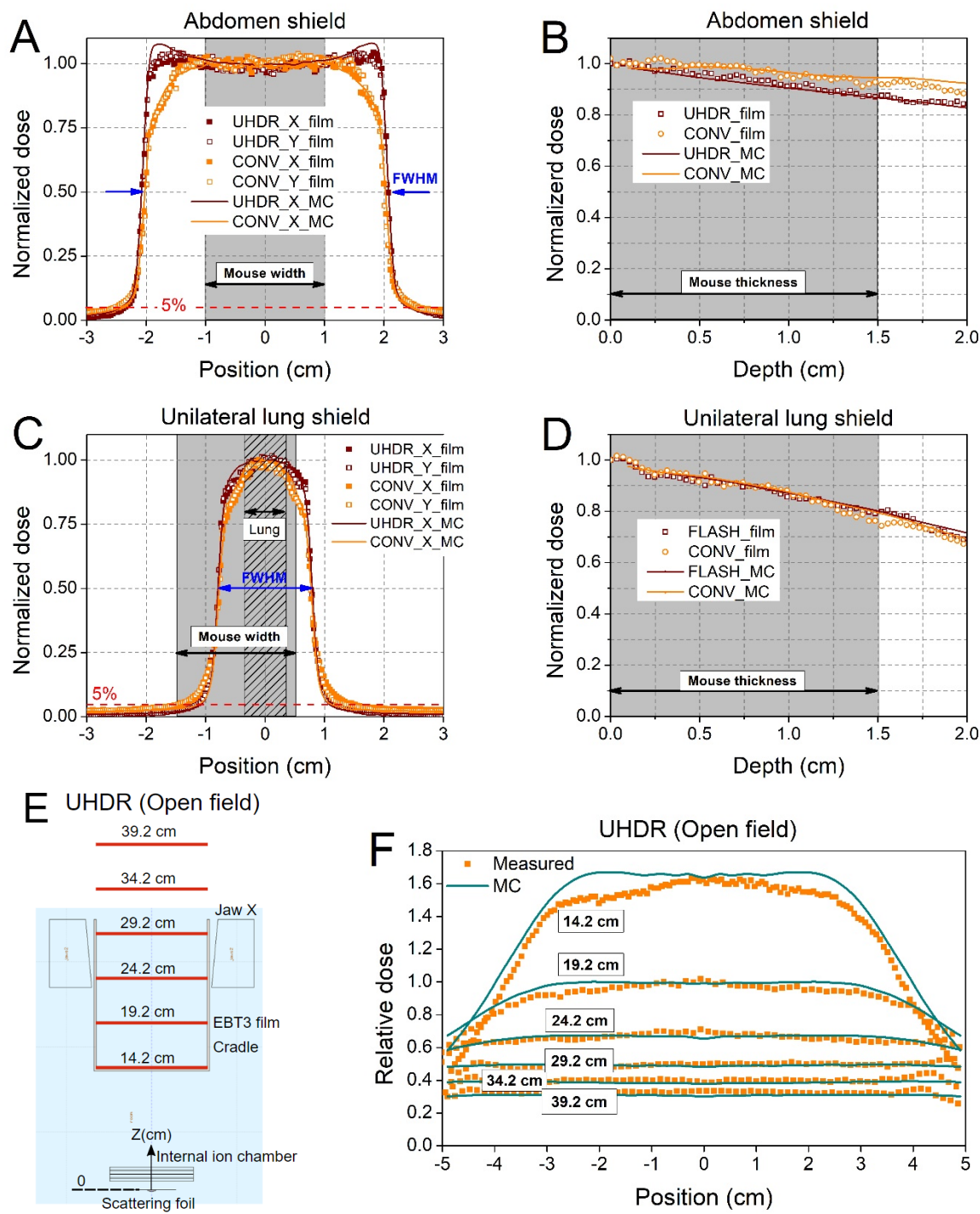

**Figure 4.** (A-D) Film-derived and Monte Carlo-calculated dose distributions. (A, C) Transverse and lateral film profiles of the (A) whole-abdomen and (C) unilateral lung mouse collimators. The films were irradiated inside the stereotactic mouse positioner, with a 2-cm solid water phantom used for backscatter. (B) Whole abdomen and (D) unilateral lung mouse collimator percentage depth dose (PDD) curves. The films were irradiated inside the stereotactic mouse positioner in a vertical position along the Z direction, sandwiched by solid water, and aligned in the center of the radiation field. (E) Schematic showing the different elevations used to characterize an open field beam. (F) Relative open field profiles at various heights away from the scattering foil, used to derive the relative flatness of the beam at each level to estimate useful working areas for cell culture studies.

The FWHM of both profiles agreed within <5% of the predefined field size for both collimators. For the whole-abdomen collimator, with its larger radiation field, the profile edge of the CONV beam was lower than that of the UHDR beam because of loss of electronic equilibrium. In the UHDR geometry, the collimator is closer to the scattering foil and the beam produces significant numbers of electrons that scatter from the inner wall because of the larger divergence angle than that for the CONV geometry, which compensates for the electronic equilibrium loss. The two profiles agree well within the width of a typical mouse (2 cm) (**Fig. 4A**; grey highlighting). In contrast, the unilateral lung collimator, with its smaller radiation field, showed minimal electron scattering, and the profiles matched in both geometries. For the whole-abdomen collimator, the flat radiation field region was wide enough to cover the width of the mice. For the unilateral lung collimator, the width of a typical mouse right lung is ~0.6 cm, and thus most of its area would be located inside the flatness of the beam (**Fig. 4C**).

When dose depth attenuation is considered, the large irradiation field of the UHDR beam appears less flat than the CONV, with mouse exit doses of 87% for UHDR *vs*. 93% for CONV (1.5-cm solid water; **Fig. 4B; Table 1**). This finding reflects the inverse square effect, which is more significant at distances closer to the source. On the other hand, in smaller irradiation fields, the loss of electron equilibrium outperforms the inverse square effect, and the PDD curves are comparable, with mouse exit dose 79% for UHDR *vs*. 76% for CONV (**Fig. 4D; Table 1**).

**Open fields for cell culture irradiation**

The geometry used to acquire film profiles for UHDR open field irradiations at different elevations from the scattering foil are shown in **Figure 4E** and **Supplemental Figure S1D**. The analyzed film profiles and the MC-simulated beam profiles in the X direction are shown in **Figure 4F** and **Supplemental Figure S1E**. The profiles were also analyzed for their 5% and 10% flatness (variation from the central dose) and relative dose rates; those results are summarized in **Table 2**. The relative dose rates were normalized to the dose rate at 19.2 cm SSD, which is the closest reference to the entrance surface dose of the mouse at the UHDR geometry.

**Table 2**. Useful field size (flatness < 5% and <10%) for cell-culture irradiations based on film measurements from open fields at different source-to-surface distances.

| SSD, cm | 5% / 10% flatness radius, cm* | Relative dose rate, Gy/s | Max dose per pulse, Gy |
|---|---|---|---|
| 14.2 | 1.2 / 2.8 | 1.60 | 4 |
| 19.2 | 2.2 / 3.3 | 1.00 | 2 |
| 24.2 | 2.5 / 4.7 | 0.69 | 1 |
| 29.2 | 3.0 / >5 | 0.50 | 1 |
| 34.2 | 3.8 / >5 | 0.39 | 1 |
| 39.2 | 4.4 / >5 | 0.33 | NA |

Abbreviations: SSD, source-to-surface distance; NA, not applicable.

* Flatness is defined as the maximum variation from the central dose; from the evaluation of both the left and right side of the measured film profiles, only the minimum radius between sides is presented.

The MC-simulated dose profile at 14.2 cm SSD had a higher deviation from all the other elevation points (relative to the measured profile from the film) because of its proximity to the scattering foil. The uniformity or flatness of the beam improved with distance from the scattering foil, but the dose rate decreased: at 24.2 cm SSD the dose rate was only 69% compared with reference (19.2 cm SSD), but its 5% flatness diameter was about 5 cm compared with 4.4 cm for reference SSD. At 29.2 cm SSD, the radiation field had half the dose rate of the reference and had a larger flat beam with a diameter of about 6 cm (5% flatness). At this SSD, the maximum achievable integer of dose per pulse is 1 Gy/pulse, and the maximum dose rate is 180 Gy/s at a repetition rate of 180 Hz, which are quite convenient for planning experimental irradiations. The 6-cm diameter of the flat beam at 29.2 cm for UHDR irradiations means that the entire well plates or large flasks cannot be irradiated, but small flasks and small dishes can be used. For CONV experiments, the cradle with the cell culture plate can be loaded inside the applicator (**Fig. 1G, Supplemental Fig. S1D**) or on top of the couch; a large flat beam can be readily obtained for both cases.

The physical parameters used in UHDR irradiation of mouse abdomen and open field cell cultures, reported as recommended in the guidelines proposed by the NRG Oncology Center for Innovation in Radiation Oncology[19], are shown in **Supplemental Table S1**.

## Discussion

For typical electron UHDR irradiations, the number of pulses required to deliver a total dose of up to 40 Gy usually ranges from 1 to 20[27], as compared with hundreds or thousands of pulses in CONV mode. Thus, failure to deliver the exact number of prescribed pulses in FLASH can produce large errors in the total delivered dose. Our initial configuration for pulse counting did not accurately measure individual pulses. Still, it acted with a time window (relay gating) of the estimated time required to deliver the desired number of pulses. In practice, parasitic capacitance in the control gating system means that the latency of the system is a distribution over time. The relay gating approach resulted in delivery of ±1 pulse to the total number of pulses required for the targeted dose; in other words, 1 in 5 total dose deliveries was missed at 90 Hz (**Fig. 3C,D**). Our optimization of the relay gating led to reduction of the dose failure rate to 1 in 20 deliveries at 90 Hz. A detailed investigation and characterization of the machine latency and a different configuration that resolves this problem have been concluded and will be presented elsewhere. For the purposes of this report, pulse delivery failures can be easily identified from the beam monitor system that uses an external ion chamber during UHDR experiments, and extra mice or samples can be irradiated to compensate for the group size of the animals or cell cultures.

The exit charge measurements of the ion chamber were proportional to the entrance dose but are subject to variation due to mouse weight or cell culture medium being used in the same type of dishes. Therefore, real-time dosimetry is possible but more complex. For example, an accurate pC/Gy can be estimated for mice from a correlation of the corrected exit charge (accounting for temperature, pressure, and mouse weight) and the film entrance dose, assuming that a sizable dataset can be collected to minimize uncertainty from the film readings. The film dose uncertainly can be up to 4% (ΔD in **Fig. 3C,D**) owing to film-to-film heterogeneity (~4%)[28,29], evolution of OD for the first 48 hours after irradiation (~ 3% change)[28], as well as heterogeneity of scanner performance and positioning of the films on the scanner[30]. The *in vivo* readings of the charges (ΔQ in **Fig. 3C**) are also subjective to the variation of the manual tuning of the automatic frequency controller. Insufficient warm-up of the machine could also result in uncertainty. The *in vivo* film readings showed a variability of 1.9% and, on average, a 1.4% dose deviation from the prescribed dose.

The average charge from the beam monitor (exit charge) showed 0.8% variability, and the average dose from the films (entrance dose) was 1.8% higher than the prescribed dose (**Fig.**

**3C**). Although these differences are not substantial, the standard practice should include passive dosimetry with films and correlations with charge readings for several experiments before relying solely on charge readings for dosimetry. Thereafter, films should be used to spot-check the calibrations. Ideally, the internal ion chamber in the treatment head could be used for real-time dosimetry by monitoring the total MU, number of delivered pulses, beam symmetry, and flatness. However, during FLASH experiments, the dose per pulse is too high, and the ion chamber saturates[11,31]. An alternative approach for real-time beam monitoring is the use of a beam-current transformer, the response of which shows a linear relationship and dose in both CONV and UHDR irradiations[32,33]. Several radiation detectors that can be used for UHDR irradiation are commercially available for evaluation, including a plastic scintillator-based dosimeter[34].

The maximum dose per pulse of our configured system was about 4 Gy/pulse when large radiation fields were to be used. Moreover, the pulse control circuit limits the frequency operation to 90 Hz (dose rate of 360 Gy/s) to avoid significant pulse failure. Thus, our configuration is not suitable for evaluating single-pulse delivery by FLASH versus CONV *in vivo*, because the dose rate would be equal to the instantaneous dose rate, i.e., on the order of millions of Gy/s. However, for investigating the FLASH effect in preclinical studies (both *in vivo* and *in vitro*), the dose rates achieved with this configuration were well above the putative FLASH effect dose-rate threshold of 40 Gy/s.

Decommissioning the 20-MeV board has been essential for this UHDR configuration, which also required a small SSD. Ideally, the SSD for the CONV configuration should match the SSD for the UHDR, but here we were limited by the clinical operation of the 16-MeV electron circuit board, which meant that we could not radically change this energy beam. The compromise was thus to use different geometries for UHDR and CONV irradiations. The energies of the two geometries at the entrance to the mouse body in our configuration were slightly different (dE ~1 MeV). The beam's edges in large radiation fields in the CONV geometry were attenuated (**Fig. 4A**; shouldering), but this attenuation can be considered in the design of the mouse collimators. The geometric difference of the beams created a small difference in depth dose attenuation in the larger (abdomen) radiation field, but overall the percentage dose at the mouse body exit was well within 90% (**Fig. 4B**). The mean profiles for the small radiation collimators were well matched, but the overall attenuation of the entrance dose at the body exit was about 75%, which should be considered when experiments are designed. Matching the UHDR and CONV doses at the center of the irradiated mouse instead of at the entrance surface can achieve dose agreement of <5% between the two setups for both collimators. In the ideal case, the same beam geometry could be used (i.e., the same SSD between modalities) if the 16-MeV beam were to be decommissioned, or if the energy of the UHDR electron beams could be further tuned to match that of the CONV beams. (These experiments are underway and will be presented in a separate publication.) CONV profiles showed higher out-of-field leakage despite the relatively lower electron beam energy of the CONV beam (**Fig. 4A,C, Table 1**), because for UHDR irradiation, scattered electrons from the sides of the wall increased the central dose and reduced the leakage. The same reason also explains the sharper penumbra in the UHDR profile compared with the CONV profile (**Fig. 4A,C, Table 1**).

For open-field UHDR irradiation, we achieved a compromise between field size and dose rate for cell culture irradiations at an SSD with a 6-cm uniform dose diameter, which was half the dose rate at the entrance point in the mouse experiments. The integer dose per pulse of 1 Gy was chosen for convenience; higher dose rates at lower elevations can be obtained by fine-tuning the dose per pulse. The MC-simulated dose profile deviated from the measured profile to a greater extent at the 14.2 cm SSD, i.e., that closest to the mirror. One possible reason for this deviation could be the symmetric electron source setup, with a 1-mm FWHM Gaussian spatial distribution, in the MC simulation, whereas in reality, the electrons may have slightly different spatial distributions. The source effect on the dose profile is weakened with distance from the source because of electron scattering. Moreover, a 16-MeV flattening filter was used for the FLASH experimental setup. However, the MC simulations indicate that the mean energy of the source electrons was 17.2 MeV (**Supplemental Fig. S2D**). Therefore, to obtain flat beams with wider diameters, the bending magnet current can be fine-tuned to select electrons with energies close to the 16-MeV beam, or the energy-specific flattening filter could be redesigned.

Both measurements and MC simulations of our 3D-printed collimators showed that radiation leakage in the mouse body region at 1 cm from the field edge was <5%, which is superior to leakage from use of a 1-cm lead collimator[18]. Our collimator design provides a fast (3D printing), cheap, hazard-free (lead is toxic), reusable ($Al_2O_3$ powder and tungsten beads), and effective solution for small animal experiments. Moreover, the thickness of the tungsten and $Al_2O_3$ layer can be further optimized to reduce leakage.

## Conclusions

This report describes technical advances and dosimetry characteristics of a Varian Trilogy clinical RT system that was reconfigured for performing UHDR experiments with mice and cell cultures. Our findings may provide insights for others planning to configure their clinical machines for UHDR research experiments, and for vendors that plan to implement electron

FLASH capabilities in their next-generation RT systems.

# References

1. Limoli C, Vozenin M-C. Reinventing Radiobiology in the Light of FLASH Radiotherapy. *Annual Review of Cancer Biology*. 2023;7.

2. Schuler E. Ultra-high dose rate electron beams and the FLASH effect: From preclinical evidence to a new radiotherapy paradigm. *Med Phys*. 2022;49:2082–2095.

3. Sørensen BS, Sitarz MK, Ankjærgaard C, *et al*. Pencil beam scanning proton FLASH maintains tumor control while normal tissue damage is reduced in a mouse model. *Radiotherapy and Oncology*. 2022;175:178–184.

4. Hughes JR, Parsons JL. FLASH Radiotherapy: Current Knowledge and Future Insights Using Proton-Beam Therapy. *Int J Mol Sci*. 2020;21.

5. Bazalova M, Nelson G, Noll JM, *et al*. Modality comparison for small animal radiotherapy: a simulation study. *Med Phys*. 2014;41:011710.

6. Rezaee M, Iordachita I, Wong JW. Ultrahigh dose-rate (FLASH) x-ray irradiator for pre-clinical laboratory research. *Phys Med Biol*. 2021;66.

7. Montay-Gruel P, Bouchet A, Jaccard M, *et al*. X-rays can trigger the FLASH effect: Ultra-high dose-rate synchrotron light source prevents normal brain injury after whole brain irradiation in mice. *Radiother Oncol*. 2018;129:582–588.

8. Darafsheh A, Hao Y, Zwart T, *et al*. Feasibility of proton FLASH irradiation using a synchrocyclotron for preclinical studies. *Med Phys*. 2020;47:4348–4355.

9. Diffenderfer ES, Verginadis, Kim MM, *et al*. Design, Implementation, and in Vivo Validation of a Novel Proton FLASH Radiation Therapy System. *Int J Radiat Oncol Biol Phys*. 2020;106:440–448.

10. Titt U, Yang M, Wang X, *et al*. Design and validation of a synchrotron proton beam line for FLASH radiotherapy preclinical research experiments. *Medical Physics*. 2022;49:497–509.

11. Poppinga D, Kranzer R, Farabolini W, *et al.* VHEE beam dosimetry at CERN Linear Electron Accelerator for Research under ultra-high dose rate conditions. *Biomed Phys Eng Express*. 2020;7.

12. Sampayan SE, Sampayan KC, Caporaso GJ, *et al.* Megavolt bremsstrahlung measurements from linear induction accelerators demonstrate possible use as a FLASH radiotherapy source to reduce acute toxicity. *Sci Rep*. 2021;11:17104.

13. Moeckli R, Goncalves Jorge P, Grilj V, *et al.* Commissioning of an ultra-high dose rate pulsed electron beam medical LINAC for FLASH RT preclinical animal experiments and future clinical human protocols. *Med Phys*. 2021;48:3134–3142.

14. Schulte R, Johnstone C, Boucher S, *et al.* Transformative Technology for FLASH Radiation Therapy. *Applied Sciences*. 2023;13.

15. Maxim PG, Tantawi SG, Loo BW. PHASER: A platform for clinical translation of FLASH cancer radiotherapy. *Radiother Oncol*. 2019;139:28–33.

16. Stephan F, Gross M, Grebinyk A, *et al.* FLASHlab@PITZ: New R&D platform with unique capabilities for electron FLASH and VHEE radiation therapy and radiation biology under preparation at PITZ. *Physica Medica*. 2022;104:174–187.

17. Wu Y (Fred), No HJ, Breitkreutz DY, *et al.* Technological Basis for Clinical Trials in FLASH Radiation Therapy: A Review. *ARO*. 2021:6–14.

18. Schuler E, Trovati S, King G, *et al.* Experimental Platform for Ultra-high Dose Rate FLASH Irradiation of Small Animals Using a Clinical Linear Accelerator. *Int J Radiat Oncol Biol Phys*. 2017;97:195–203.

19. Zou W, Zhang R, Schüler E, *et al.* Framework for Quality Assurance of Ultrahigh Dose Rate Clinical Trials Investigating FLASH Effects and Current Technology Gaps. *International Journal of Radiation Oncology*Biology*Physics*. 2023;116:1202–1217.

20. Felici G, Barca P, Barone S, *et al.* Transforming an IORT Linac Into a FLASH Research Machine: Procedure and Dosimetric Characterization. *Frontiers in Physics*. 2020;8.

21. Soto LA, Casey KM, Wang J, *et al.* FLASH Irradiation Results in Reduced Severe Skin Toxicity Compared to Conventional-Dose-Rate Irradiation. *Radiat Res*. 2020;194:618–624.

22. Levy K, Natarajan S, Wang J, *et al.* Abdominal FLASH irradiation reduces radiation-induced gastrointestinal toxicity for the treatment of ovarian cancer in mice. *Sci Rep*. 2020;10:21600.

23. Khan S, Bassenne M, Wang J, *et al.* Multicellular Spheroids as In Vitro Models of Oxygen Depletion During FLASH Irradiation. *Int J Radiat Oncol Biol Phys*. 2021;110:833–844.

24. Simmons DA, Lartey FM, Schüler E, *et al.* Reduced cognitive deficits after FLASH irradiation of whole mouse brain are associated with less hippocampal dendritic spine loss and neuroinflammation. *Radiother Oncol*. 2019;139:4–10.

25. No HJ, Wu Y (Fred), Dworkin ML, *et al.* Clinical Linear Accelerator–Based Electron FLASH: Pathway for Practical Translation to FLASH Clinical Trials. *International Journal of Radiation Oncology\*Biology\*Physics*. 2023:S0360301623003681.

26. Battistoni G, Bauer J, Boehlen TT, *et al.* The FLUKA Code: An Accurate Simulation Tool for Particle Therapy. *Front Oncol*. 2016;6:116.

27. Bourhis J, Montay-Gruel P, Goncalves Jorge P, *et al.* Clinical translation of FLASH radiotherapy: Why and how? *Radiother Oncol*. 2019;139:11–17.

28. Sipila P, Ojala J, Kaijaluoto S, *et al.* Gafchromic EBT3 film dosimetry in electron beams - energy dependence and improved film read-out. *J Appl Clin Med Phys*. 2016;17:360–373.

29. Jaccard M, Petersson K, Buchillier T, *et al.* High dose-per-pulse electron beam dosimetry: Usability and dose-rate independence of EBT3 Gafchromic films. *Med Phys*. 2017;44:725–735.

30. Lewis D, Micke A, Yu X, *et al.* An efficient protocol for radiochromic film dosimetry combining calibration and measurement in a single scan. *Med Phys*. 2012;39:6339–50.

31. Di Martino F, Giannelli M, Traino AC, *et al.* Ion recombination correction for very high dose-per-pulse high-energy electron beams. *Med Phys*. 2005;32:2204–10.
32

32. Oesterle R, Gonçalves Jorge P, Grilj V, *et al.* Implementation and validation of a beam-current transformer on a medical pulsed electron beam LINAC for FLASH-RT beam monitoring. *J Appl Clin Med Phys*. 2021;22:165–171.

33. Liu K, Palmiero A, Chopra N, *et al.* Dual beam-current transformer design for monitoring and reporting of electron ultra-high dose rate (FLASH) beam parameters. *J Appl Clin Med Phys*. 2023;24:e13891.

34. Ashraf MR, Rahman M, Cao X, *et al.* Individual pulse monitoring and dose control system for pre-clinical implementation of FLASH-RT. *Phys. Med. Biol.* 2022;67:095003.

# Supplemental Material

**Operation**

For UHDR irradiation, the machine is operated in service mode with the decommissioned 20 MeV energy board, with the dose servos turned OFF, the motor pot ON, and the auto frequency controller switched to manual. To bypass clinical checkpoints, several interlocks must be overwritten (DOS2, DS12, XDP1, XDP2, XDRS, XDR1, XDR2, EXQ1, EXQ2, EXQT, DOS1, TIME, DPSN, MOTN, COLL, IPSN, ACC, TDRV, GFIL, and MLC). The desired number of pulses, frequency, and latency are controlled by a computer connected wirelessly to the pulse control circuit; C code is delivered after the beam is turned ON to ensure that the voltage of the feeding pulse wave is at maximum. Linac operations for the CONV irradiations involve use of the clinical 16-MeV energy board, and the dose is estimated and delivered in monitor units (MU). However, the machine is also operated in service mode because the system does not allow the delivery of a large number of MU when run in clinical mode. Furthermore, the reduced SSD in CONV mode also requires that the MUs be delivered at 72 Hz (repetition rate 4) to match the mean dose rate used clinically (which is delivered at 180 Hz).

**Pulse control and pulse counting**

A schematic of the electronic circuit that controls the pulses during UHDR irradiations is shown in **Figure 1A**. The dose per pulse is controlled by the gun current (Gun I), which was modified from that of the 20-MeV circuit board. To avoid gradual ramp-up of the intensity of the first few pulses and to ensure that all pulses will be delivered at maximum intensity, the auto frequency controller was set to manual control and adjusted to achieve resonance as monitor through an oscilloscope of the reflected and feeding wave in the linac[1]. The microcontroller (Red Pitaya, Slovenia, Europe) reads the voltage signal from the top ionization chamber (Top TP1) to control the number of delivered pulses. A relay-gating time window is executed via compiled C code on a personal computer connected wirelessly to the microcontroller, to control the number of pulses. The machine latency (resulting from the need to check for the beam hold request) leads to ± 1 pulse being delivered with a failure frequency of 1 in 5 deliveries (i.e., 20%) at a repetition rate of 90 Hz. Thereafter, the latency time during beam ON and beam OFF distributions can be characterized to reduce uncertainty. To aim for certain number of pulses the opening the relay was set at 1.3 ms before the last intended pulse to stop pulse delivery. With this approach, the first pulse after the relay is opened is delivered due to machine latency, and the second pulse cannot be delivered because it is out of the latency range (**Fig. 1B**). According to this pulse-counting algorithm, the frequency of an extra pulse being delivered is reduced to 1 in 20 pulses at 90 Hz.

**Collimator design and production**

The well plate holder for cell irradiation, the mouse irradiation collimators, with various exposure field apertures for anatomically specific irradiations (**Supplemental Fig. S1A**), and the stereotactic mouse positioners were designed with Fusion 360 (Autodesk, San Rafael, CA) and printed by Ultimaker S5 (Ultimaker, Utrecht, Netherlands) from polylactic acid plastic.

**Stereotactic mouse positioner**

The stereotactic mouse positioner used for whole-abdomen irradiation of mice under ketamine anesthesia (shown in **Fig. 1C**) has inner dimensions of $2.7 \times 10.7 \times 2.6$ cm$^3$ (W × L × H) and has 0.2-cm-thick walls except for the area to be exposed to radiation, where the bottom thickness is reduced to 0.1 cm; thus the area for loading the radiochromic film for dosimetry is 2.7 × 5.5 cm$^2$. To accurately position the mouse, a metal wire is mounted at the front of the positioner to serve as a teeth bar, and the mouse's tail is passed through a rear hole and taped to the exterior of the lateral wall, with a small amount of traction applied for immobilization and reproducible positioning. The extruded ridges on the exterior sides of the positioner (front lateral sides and back) also serve as positioning anchors to allow precise placement of the mouse positioner on top of the collimator (**Fig. 1D**). Every such collimator has a matching stereotactic mouse positioner with the appropriate anchoring, adjusted film loading area, and marks on the exterior walls to indicate the area to be irradiated.

**Anatomically specific mouse collimators**

The collimator used for whole-abdomen irradiation of mice (field size 4×4 cm$^2$) is designed for radiation to be directed from the bottom to the top of the collimator (**Fig. 1D**); it is printed as a hollow shell with dimensions of 8×12×4 cm$^3$ (W×L×H). The collimator has a horizontal separator that splits its volume into one 3-cm bottom layer and one 1-cm top layer. The bottom compartment is filled with a relatively low-Z aluminum oxide powder (11067 Al$_2$O$_3$, 325 mesh 99.5%, Alfa Aeser, MA) that slows the primary incoming electrons while producing less bremsstrahlung. The top layer is filled with 2-mm tungsten spheres (W, THPP, San Diego, CA) with nominal density of 17.5 g/cm$^3$ to absorb the bremsstrahlung X-rays generated from the Al$_2$O$_3$. The collimator's inner walls, which shape the radiation field, are printed with a thickness of 0.04 cm (a value limited by the printer's resolution). The extruded ridges on the top of the collimator serve as positioning anchors for the precise placement of the stereotactic mouse positioner.

**Cradle-collimator positioning**

Achieving high-dose pulses requires precise positioning of the mouse collimator inside the treatment head; to achieve this, we machined a custom cradle from transparent polystyrene (**Fig. 1E**). The lateral dimensions of the cradle (10.7×13.6 cm$^2$) were limited by the maximum jaw opening inside the treatment head, and its height (17.6 cm) by the depth of the light field mirror inside the treatment head (**Fig. 1F).** The top of the cradle is designed to fit precisely on the surface of the linac's treatment head, below the multileaf collimator (**Fig. 1G**, see UHDR). For the CONV geometry, the cradle is placed inside the 15×15 cm$^2$ applicator and accurately positioned by using clinical lasers (**Fig. 1G**; CONV). One side of the cradle is open, allowing the collimator to slide outside the cradle at a predetermined distance, which is used to accurately align the center of the collimator's irradiation area with the center of the electron beam (**Fig. 1E**; **Supplemental Fig. S1B,C**). The mouse irradiation field can be accurately aligned at the center of the beam by using the appropriate slider top for the 'anatomic collimators,' which allow irradiation of different regions of the mouse (**Supplemental Fig. S1B,C**; of the various collimators, the brain collimator requires the maximum displacement for alignment). The lateral dimensions of the cradle are wide enough to fit a 10-cm tissue culture dish or well plates for cell irradiation. For irradiation of cell cultures, a cell-designated holder is mounted on the top of the cradle (**Supplemental Fig. S1B,C**). Only one of the two devices (cell culture positioner or anatomic collimator) is mounted on the cradle at any one time.

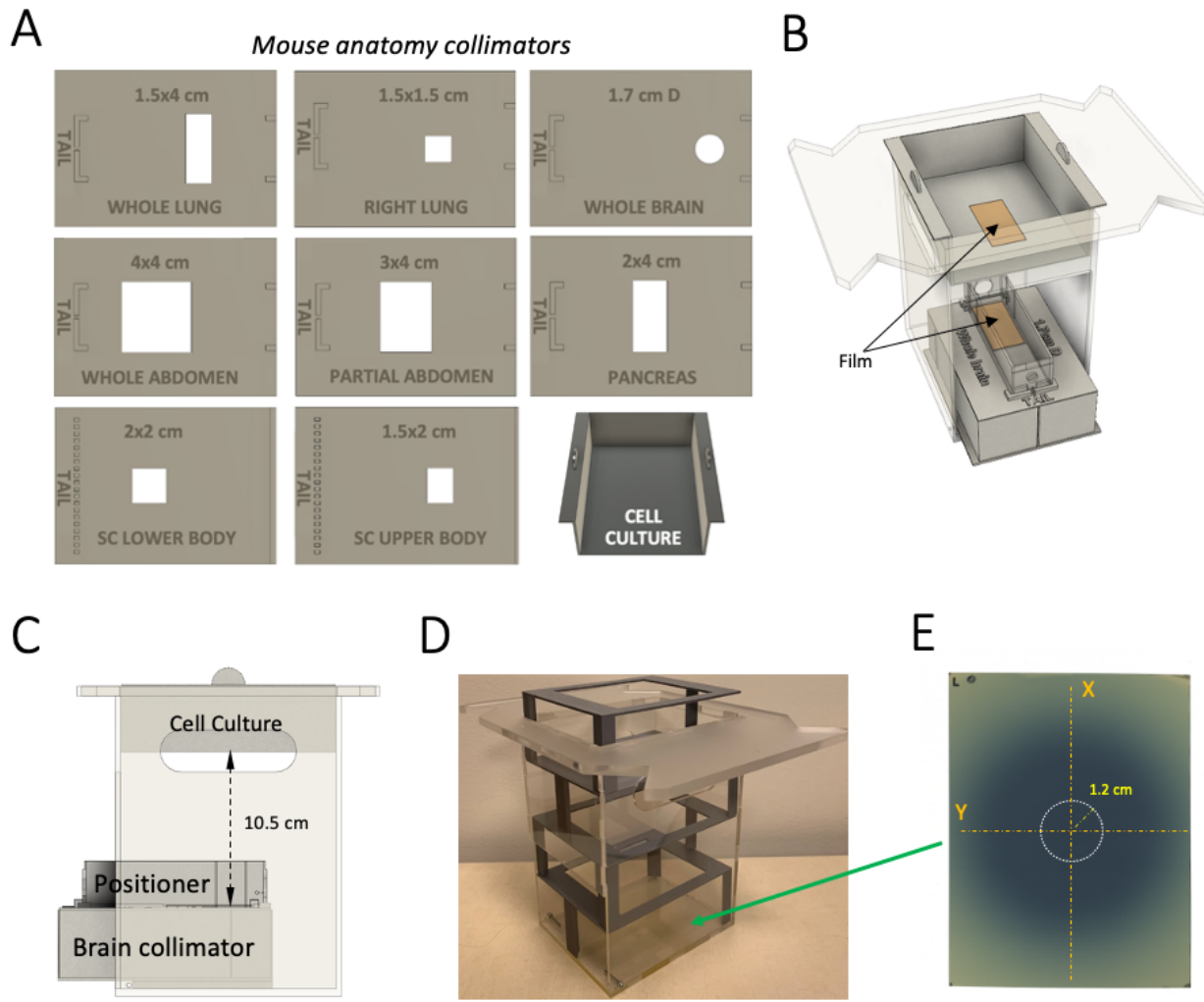

**Supplemental Figure S1**. (A) 3D CAD files of the various designs for mouse irradiation collimators used for in vivo organ-specific exposures and from the cell culture positioner for well-plates, flasks, or tissue culture dishes. (B) A 3D CAD file shows the cradle that suspends the mouse collimator with the stereotactic mouse positioner or cell culture positioner at a specific source-to-surface distance (SSD). Although the beam collimator and cell culture holder are shown with both loaded on the cradle for illustration, in practice only one at the time would be loaded for a study. The positioning of the radiochromic film used to measure the entrance dose on the irradiated subject is also shown. (C) The film-to-film distance from the animal collimator to the cell culture positioner is 10.5 cm, which lengthens the SSD of cell culture work by 10.5 cm. (D) Photograph showing the 3D printed structure used for the suspension of the radiochromic films for evaluating the beam flatness of an open field at different distances for the scattering foil (14.2, 19.2, 24.2, 29.2, 34, 2 and 39.2 cm). (E) Scan of a 10 × 13 cm radiochromic film irradiated with UHDR open field at 14.2 cm from the scattering foil. The green arrow indicates the placement of the film inside the cradle. The shape of the open field is circular, and the dotted

circle indicates 5% flatness of the field in this SSD, which has minimal coverage for cell culture work. The dotted lines indicate the analyses of the profiles in the Z and Y direction for determining the flatness of the field.

**Beam monitor for *in vivo* UHDR irradiation**

A PWT Farmer Chamber 30013 (Booton, NJ) was used to measure the beam charges. The sensitivity of the ion chamber requires that the electron beam be attenuated to avoid saturation. The bremsstrahlung from an attenuated electron beam is directly proportional to the number of electrons in the beam. The bremsstrahlung tail is an ideal energy deposition region for the sensitivity of the Farmer chamber[2]. Solid water beam attenuation is typically used in clinical measurements with Farmer chambers, in which the chamber is placed inside a Farmer-suitable 2-cm-thick block of solid water with a borehole in its center. We established the thickness of the solid water before the chamber and the positioning of the Farmer chamber in relation to the radiation source by determining the bremsstrahlung tail by using solid water on the top of the couch. The couch can be raised 20 cm from the isocenter (VRT +20 cm) before triggering collision interlock of the system. A PDD curve was derived from a large piece of film (20.3×25.4 cm$^2$) to establish the thickness of solid water required for the attenuation. The film was positioned vertically within slabs of solid water, parallel along the Z direction, on the top of the couch, for both modalities (**Supplemental Fig. S2A**). A 10-cm solid water build-up was found to be sufficient for attenuating the electrons from the beam, with a projected range of 9 cm measured from the film-derived PDD (**Supplemental Fig. S2B**). The Farmer-suitable 2-cm solid water block with the ion chamber was placed on top of the 10-cm solid water build-up, and an additional 3-cm solid water block was placed above the ion chamber for backscatter (**Fig. 1G**). To obtain information on field homogeneity (flatness) and to evaluate setup uncertainty at the height at which the ion chamber would be located, a large piece of film (20.3×25.4 cm$^2$) was irradiated flat on the top of an 11.0-cm solid water block with a 3-cm block above the film for backscatter, and lateral profiles were derived along X and Y directions for both modalities (**Supplemental Fig. S2A,C**). The ion chamber readings are monitored from an electrometer (MAX 4000 PLUS SN: J162212, Middleton, WI) located in the control room.

**Film analysis**

Each film batch is calibrated to determine the net optical density (OD) versus dose gradient and scanned at different times after irradiation according to the manufacturer's film calibration protocol. These experimental films were scanned at 24 hours after irradiation by using an Expression 10000XL Photo scanner (Epson, Los Alamitos, CA). The films were placed precisely in the middle of the scanner and scanned at positive 48 bits color, with no color correction, at 72 dots-per-inch resolution and saved as tiff files (Epson Scan). To calculate the dose from OD, we used a ratio suggested by the film vendor: Dose, in Gy = a / (OD-b) + c, where a, b, and c are fitting parameters. The net optical densities for all three-color channels were processed by using MATLAB (MathWorks Inc.), and the average of a large area within the irradiated area was assessed. The green channel was estimated to have a steeper OD than the dose gradient in the higher dose region and was used for calculating dose to the experimental films.

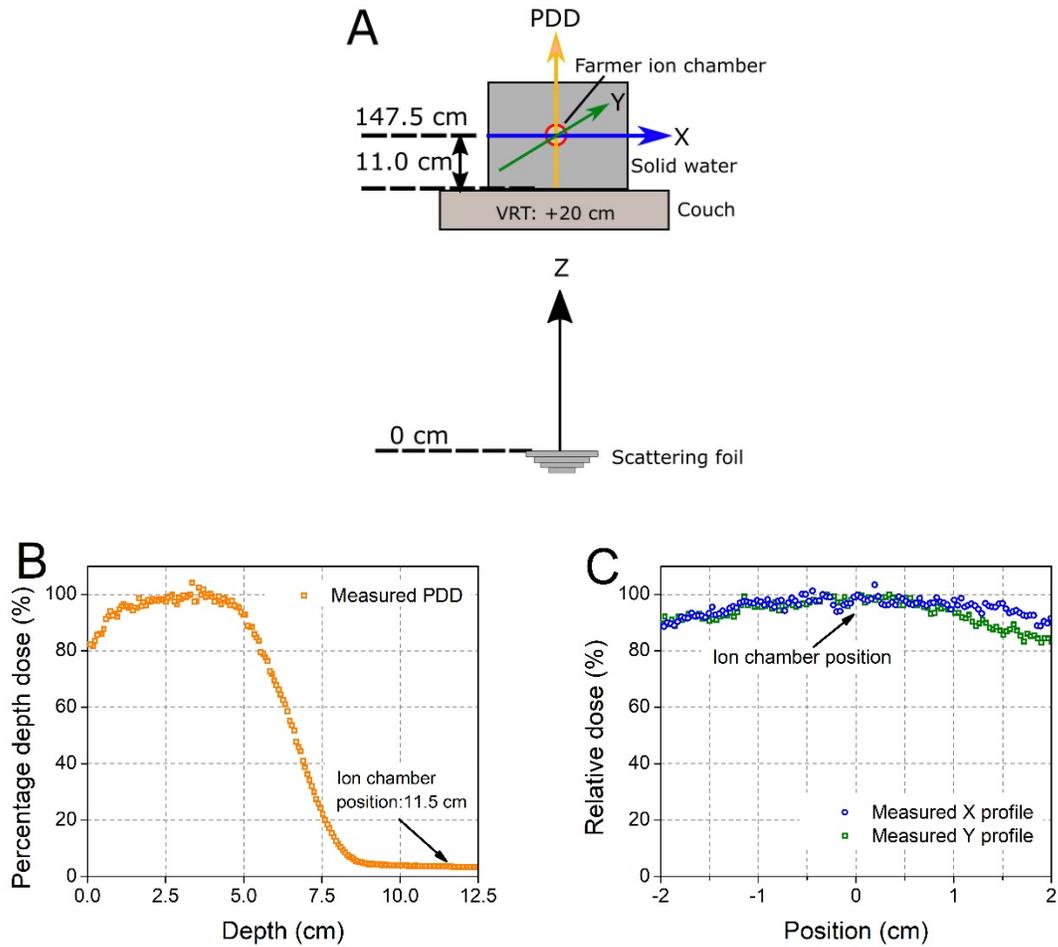

**Supplemental Figure S2**. (A) Experimental setup for measuring percentage depth dose (PDD) and transverse profiles on top of the couch. (B) Percentage depth dose curve. The ion chamber is placed in the bremsstrahlung tail to avoid chamber saturation. (C) Dose profiles along the x- and y-axes. The central region is relatively flat, so chamber measurement is insensitive to the lateral position uncertainty.

**Supplemental Table S1**. Typical physical parameters for electron UHDR irradiations.

| Parameters | Abdomen collimator | Open field at SSD 29.2 cm |
|---|---|---|
| Mean electron beam energy, MeV | 17.2 | 17.2 |
| Pulse frequency, Hz | 90 | 180 |
| Pulse width, µs | 3.75 | 3.75 |
| Dose per pulse, Gy | 2 | 1 |
| Mean dose rate, Gy/s | 210 | 180 |
| Instantaneous dose rate, Gy/s | 5.3E+5 | 2.7E+5 |
| Beam field size | Square field (4×4 cm$^2$) | Circular field (flatness 5%) (6-cm diameter) |

Abbreviations: UHDR, ultrahigh dose rate; SSD, source-to-surface

# References

1. Schuler E, Trovati S, King G, *et al.* Experimental Platform for Ultra-high Dose Rate FLASH Irradiation of Small Animals Using a Clinical Linear Accelerator. *Int J Radiat Oncol Biol Phys*. 2017;97:195–203.

2. Arunkumar T, Supe SS, Ravikumar M, *et al.* Electron beam characteristics at extended source-to-surface distances for irregular cut-outs. *J Med Phys*. 2010;35:207–214.